# Apache Web Server Execution Tracing Using Third Eye

**Raimondas Lencevicius**   **Alexander Ran**   **Rahav Yairi**

Nokia Research Center, 5 Wayside Road, Burlington, MA 01803, USA

Raimondas.Lencevicius@nokia.com   Alexander.Ran@nokia.com   Rahav.Yairi@nokia.com

**ABSTRACT**

Testing of modern software systems that integrate many components developed by different teams is a difficult task. Third Eye is a framework for tracing and validating software systems using application domain events. We use formal descriptions of the constraints between events to identify violations in execution traces. Third Eye is a flexible and modular framework that can be used in different products. We present the validation of the Apache Web Server access policy implementation. The results indicate that our tool is a helpful addition to software development infrastructure.

## 1  INTRODUCTION

Nowadays, many software-intensive systems such as personal communication devices or communication network elements integrate many dozens of software components that are designed to run on different types of hardware, to interoperate with different environments and to be configurable for different modes of operation and styles of use. The situation is further complicated by a fact that these components are often developed by geographically distributed teams, using different programming languages, development tools and even different design and development methodologies. All this makes complete testing of these systems in a lab very hard.

In the Third Eye project [4], we focused on abstract representation of software - its architecture. We have defined a methodology for tracing execution of software by reporting events meaningful in the application domain or essential from the implementation point of view. Many ideas incorporated in the Third Eye framework were inspired by the Logic Assurance system [5] and work on enforcing architectural constraints [2]. In Third Eye, we have used different technologies to make the framework more extensible, to allow its integration with other trace analysis tools and specification languages. The implemented portable prototype of the Third Eye framework includes reusable software components for event definition and reporting and stand-alone tools for storage and query of event traces, constraint specification and trace analysis.

## 2  THIRD EYE ARCHITECTURE

A central decision of Third Eye framework is what information from the execution state of the program is traced. We decided to trace occurrences of *events* (Figure 1). Events cross the boundary between the application and implementation domains allowing abstract specifications that use event properties and a simple representation in the implementation domain. Such representation helps to produce traces without introducing new errors. "Event" in this case is a qualitative change in the state of an entity either meaningful in the application domain or significant architecturally. In Third Eye framework, events are typed objects. Event type has a name, a list of named and typed properties, and the type constructor. Third Eye event types are similar to classes in programming languages although the only method associated with the event type is its constructor. We allow event type inheritance. To report an event, developers specify the type of the event and set values of the event properties. Developers need to set only the properties that were not set already by the event constructor. Events in Third Eye are characterized by the time and location of their occurrence.

Correct behavior specifications define constraints on the properties of the events, their sequence, location, and timing. We use formal descriptions of the constraints between events to identify violations in execution traces.

Another concept of Third Eye is the *tracing state*. A tracing state is a set of event types generated in that state. Other event types are filtered out and not reported. The system is always in a certain tracing state. Tracing states correspond to specifications. A program specification describes a set of constraints on events. The event types used in a specification have to be monitored to validate a trace against this specification. All event types contained in a specification and monitored for this specification form a tracing state. Tracing states also control the overhead of tracing on the executing system.

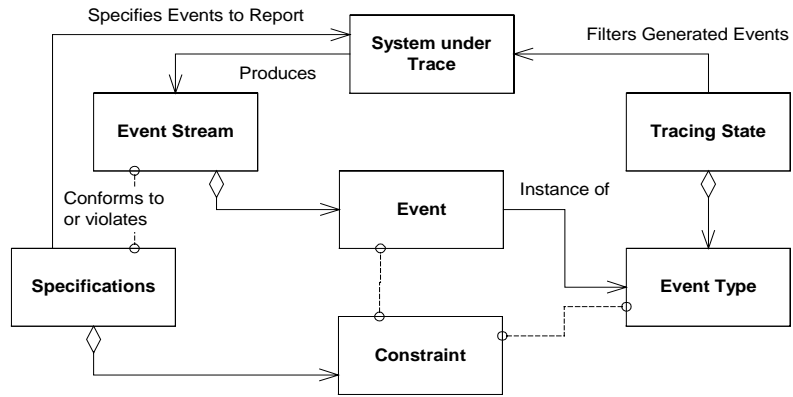

**Figure 1. Third Eye Conceptual Architecture**

The Third Eye framework includes modules for event type definition, event generation and reporting, tracing state definition and management, trace logging, query and browsing interfaces (Figure 2). Modules of event type definition, event reporting facility and tracing state controler are integrated with the software of the system under trace (SUT). The rest of the modules are independent from the SUT and can be deployed on a different execution platform to minimize the influence on system performance.

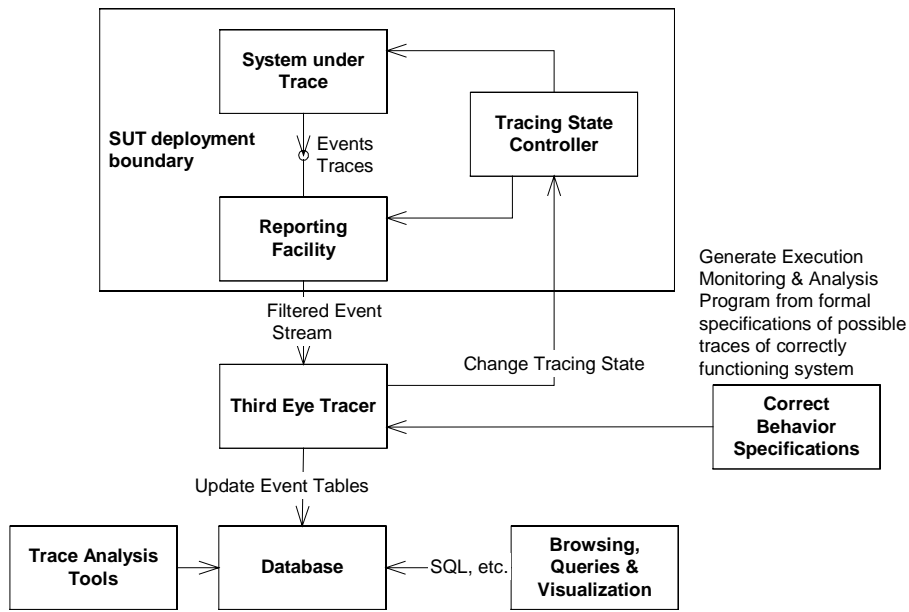

**Figure 2. Third Eye Module Structure**

The module structure was partitioned along the lines of standard interfaces to achieve portability and to enable integration to third party software and tools. For example, the event reporter and tracer can be connected through a file, a socket, or using an ORB. Trace delivery for logging and analysis uses alternative interfaces to accommodate devices with different data storage and connectivity capabilities.

We have implemented a Third Eye framework prototype. This prototype is currently used by the Third Eye project team in collaboration with product development teams in Nokia's business units. We used Third Eye to test a number of software systems: the memory subsystem of one of Nokia's handsets, Apache Web Server, WAP (Wireless Application Protocol) client, and Internet browser. Our experience shows that architecturally significant events can be identified and constraints among them can be established and checked using our framework. The rest of this paper describes Apache Web Server tracing and trace analysis.

## 3  APACHE WEB SERVER ACCESS POLICY TESTING

In this paper we describe the correctness analysis of the Apache Web Server access policy using the Third Eye framework.

### 3.1  Apache Access Policy

The Apache Web Server access policy is described in "Apache: The Definitive Guide" [3]. The user's access rights to web pages served by Apache depend on the user's domain and on the location—directory and file—of the accessed page. The following summarizes the access policy.

The access rules are specified for a file or a set of files in a directory. We call this set of files a *location*. We consider three configuration directives:

- `Allow from host {host}` – this directive allows access to the location from hosts specified in the directive. Host specifications may contain wildcards.

- `Deny from host {host}` – this directive denies access to the location from hosts specified in the directive. Host specifications may contain wildcards.

- `Order ordering` – this directive specifies in which order `allow` and `deny` directives for a location should be processed. There may be more than one `allow` and `deny` directive per location. By default, the ordering is "`deny,allow`". This means that `deny` directives are processed first, and `allow` directives next. The ordering can also be "`allow,deny`" or "`mutual-failure`" (which is beyond the scope of this example). The "`deny,allow`" ordering allows access from all hosts not mentioned in any allow or deny directives, while the "`allow,deny`" ordering denies access from all such hosts.

Though the meaning of directives above is not completely intuitive, they provide a powerful way for a Web server to specify complex rules for access to a location. Consider the following configuration directives:

```
Order allow,deny
Allow from goodguys.org
Deny from badguys.com
Allow from 127.0.0.1 // Localhost
Allow from 123.156.3.5
```

This configuration allows access to the location only from `goodguys.org`, `127.0.0.1` (`localhost` – the server computer itself), and IP address `123.156.3.5`. Access from other sites is denied, because they are not mentioned in any directives. If `badguys.com` maps to the IP address `123.156.3.5`, the access from `123.156.3.5` will be denied because the `deny` directives are processed after the `allow` directives according to the ordering. Formal rules for the access algorithm are given in the next section.

## 3.2 Apache Access Policy Testing

To check the correctness of the Apache access policy implementation, we instrumented the Apache Web server [1] to report Third Eye events and then checked the event trace using a policy specification written in Microsoft Access database[1]. We defined the following event types:

- `Access_order` – reports ordering directive for a certain location. Event properties: location and ordering. These events occur when the server reads a configuration file and processes access directives contained in it.

- `Access_allow` – reports allow or deny directive for a location with host names specified in the directive. Properties: location, host, and `allow/deny` directive. These events also occur when the server reads a configuration file.

- `Access_request` – reports a request of a location by a host. The event records the access code indicating whether the request was granted or denied. It also records the accessed location, its URI, and the request record. Apache uses zero access code for granted access and non-zero error code for denied access. Properties: location, request, access code and URI. These events occur when clients request web pages from the Web server.

All three event types by default also include timestamp property that allows to order directives by the time they were read from the configuration files and to order access requests by their execution time.

The running instrumented Apache server reports the Third Eye events to the Third Eye tracer, which then places them into a database. We wrote a specification of the Apache access policy in the database using a mixture of the database 3GL GUI and plain SQL. Here we outline the specification.

First, we determine the ordering of a location. We group all locations into sets according to their ordering. After that we find all requests that should be allowed access or denied access according to the configuration. We use the following rules derived from the access policy:

- If the ordering is "`allow,deny`",
  A. If there is an `allow` directive for a host and no `deny` directive for that host, the request should be allowed.
  B. If there is no `allow` directive for a host or a `deny` directive exists for the host, the request should be denied.

- If the ordering is "`deny,allow`".
  A. If there is no `deny` directive for a host or an `allow` directive exists for the host, the request should be allowed.
  B. If there is a `deny` directive for a host and no `allow` directive, the request should be denied.

Finally, we examine the access codes received by the access request to determine whether the server gave correct access rights. I.e., if the access should have been allowed by the configuration, but was denied by the server, this action was incorrect. When we ran the queries on a real-life trace recorded by an Apache Web Server, we found no errors. We introduced some errors in the Apache access policy implementation, and if these errors resulted in incorrect accesses, they were found by our trace analysis. As noted in trace analysis literature, there is no guarantee against false positives or false negatives. In a false positive, analysis may not uncover existing errors, if the test cases do not exercise the incorrectly implemented part of constraint. For example, if access would be incorrectly granted only when there is "`deny,allow`" ordering and a `deny` directive for a host, but no `allow` directive, the test case would need to contain exactly these events to catch the error. This indicates need for good test case selection. False positive may also occur when the error does not break given constraints. For example, given constraints provide no guarantees against memory leaks in the access module. Other events and rules would be needed to uncover memory leak errors. Finally, a false negative may occur if analysis incorrectly shows existence of an error, for example, if the constraints or events incorrectly reflect the desired program state.

---

[1] Microsoft database is named "Access" and should not be confused with the Apache access module or policy. To prevent such confusion, we refer to it as a "*database*".

### 3.3 Discussion

Microsoft Access database was used to store the reported events and to specify policy constraints. The database provides a user-friendly interface to the tables and supports quick report generation that is useful for simple event statistics. However, writing SQL queries is not a simple task. Microsoft Access provides a graphical query design tool that simplifies query creation, but is not adequate for complicated queries. Union queries have to be entered by hand in SQL. Overall, the job may be accomplished in a simpler way using a Prolog fact database and Prolog rules. Since Third Eye is a flexible framework that supports different plug-in components, we stored events in both an SQL database and a Prolog fact database. Consequently users can get both prefabricated database reports and write simpler Prolog rules for constraints. We are planning to further explore different back ends for trace analysis and their user interfaces.

## 4 CONCLUSIONS

Third Eye can be used for debugging, monitoring, specification validation, and performance measurements. These scenarios use typed events—a concept simple and yet expressive enough to be shared by product designers and developers. The Third Eye has an open architecture allowing easy replacement of third-party tools, including databases, analysis and validation tools. We have applied Third Eye to test a number of software systems: the memory subsystem of one of Nokia's handsets, Apache Web Server, and WAP (Wireless Application Protocol) client. Our experience shows that the Third Eye is a practical framework for specification-based analysis and adaptive execution tracing of software systems.